\documentclass[12pt,a4paper]{article}

\DeclareFontFamily{OT1}{rsfs}{}
\DeclareFontShape{OT1}{rsfs}{m}{n}{ <-7> rsfs5 <7-10> rsfs7 <10->rsfs10}{} \DeclareMathAlphabet{\mycal}{OT1}{rsfs}{m}{n}

\newcommand{\be}{\begin{equation}}
\newcommand{\ee}{\end{equation}}
\newcommand{\beq}{\begin{eqnarray}}
\newcommand{\eeq}{\end{eqnarray}}

\usepackage{amsmath,amsthm,amsfonts}
\input amssym.def
\newsymbol\square 1003
\begin{document}

\centerline{\bf  GENERALIZATION OF THE GROSS-PERRY METRICS}

\bigskip
\centerline{M. Jakimowicz and J. Tafel}

\null

\centerline{Institute of Theoretical Physics, University of Warsaw,}
\centerline{Ho\.za 69, 00-681 Warsaw, Poland, email: tafel@fuw.edu.pl}

\bigskip
\bigskip
\begin{abstract}
A class of $SO(n+1)$ symmetric solutions of the (N+n+1)-dimensional Einstein equations is found.  It contains 5-dimensional metrics of Gross and Perry and Millward.
\end{abstract}

\section{Introduction}

An extra spatial dimension was  indroduced by Kaluza and Klein (see e.g. \cite{CJ}) in order to unify electromagnetism and gravity. Recently, due to the string theory, extra dimensions became a permanent part of the high energy theoretical physics. For instance, in brane-world models (see \cite{M} for a review) matter fields are confined to a four-dimensional brane and gravity can propagate in higher dimensional bulk. These higher dimensional models motivate  studying Einstein's equations in $D>4$ dimensions. Some techniques of the  4-dimensional Einstein theory were already generalized to higher dimensions. They refer mainly to the classification of the Weyl tensor , the Robinson-Trautman solutions , spacetimes with vanishing invariants  and metrics with D-2 abelian symmetries (see e.g. \cite{CMPP,PO,CFHP,H}). 

Symmetry assumptions  are one the most efficient methods of solving the Einstein equations. All well known multidimensional exact solutions like the Myers-Perry black hole \cite{MP}, black ring of Emparan and Reall \cite{ER} and the Gross-Perry metrics \cite{GP} (see also \cite{DO}) admit several dimensional symmetry groups.

In \cite{1} we proposed a construction of vacuum metrics admitting $SO(n+1)$
spherical symmetry, which was based on the symmetry reduction of $(N+n+1)$--dimensional Einstein equations to $(N+1)$--dimensional equations with a scalar field $\phi$. There was used an additional assumption that the field of normal vectors to surfaces $\phi=const$ is geodetic and the induced metric of the surfaces is an Einstein metric. The construction, for zero cosmological constant and timelike surfaces $\phi=const$, can be summarized as follows.

Let $\gamma_{ij}$ and $P_{ij}$ be symmetric tensors depending on
coordinates $x^i$, $i=0,\ldots$,$N-1$. Assume that $\gamma_{ij}$ has
the Lorentz signature and $P_{ij}$ satisfies the 
following conditions 
\be
\ \ P^i_{\ i}=0,\label{3.17a}
\ee
\be
P^i_{\ j}P^j_{\ i}=2c=const,\label{3.18b}
\ee
\be
P^k_{\ i;k}=0\ ,\label{3.25}
\ee
where $P^i_{\ j}=\gamma^{ik}P_{ij}$ and the semicolon denotes covariant
derivative related to the metric $\gamma_{ij}dx^idx^j$. From matrices
$\gamma=(\gamma_{ij})$ and $P=(P^i_{\ j})$ we compose metric corresponding to
$\gamma e^{P\tau}$, where $\tau$ is a function of another coordinate $s$. We
assume that its  Ricci tensor satisfies
\be 
R^i_{\ j}(\gamma e^{P\tau})=\lambda\delta^i_{\ j}\ ,\ \ \lambda=const\ .\label{3.24a}
\ee
Given $c$ and $\lambda$ we look for solutions $\beta(s)$ and $\phi(s)$ of the
following equations
\be
(\beta \dot \phi)\dot{\phantom{l}}=-\beta V_{,\phi}\ ,\label{3.14}
\ee
\be 
-N\lambda\beta^{-2/N}=(1-\frac{1}{N})\frac{\dot\beta^2}{\beta^2}-\frac{2c}{\beta^2}
-\dot\phi^2+2V\ ,\label{3.23}
\ee
where the dot denotes the partial derivative with respect to s and $V$
is a function of $\phi$.

The main result of \cite{1} is that, under conditions (\ref{3.17a})-(\ref{3.23}), metric 
\be
\tilde g=-ds^2+\tilde g_{ij}dx^idx^j,\label{3.4}
\ee
were
\be
\tilde g_{ij}=\beta^{2/N}(\gamma e^{P\tau})_{ij}\label{3.20}
\ee 
and $\tau(s)$ is defined via equation
\be
\beta\dot\tau=2\ ,\label{3.21}
\ee
satisfies $(N+1)$--dimensional Einstein equations with the scalar field $\phi$
and potential $V$. Moreover, if
\begin{equation}\label{21}
V=-\frac{1}{2}n\left(n-1\right)e^{-2\sqrt{\frac{n+N-1}{n\left(N-1\right)}}\phi}
\end{equation} 
then
\be
g=e^{-2\sqrt{\frac{n}{(N-1)(n+N-1)}}\phi}\tilde g-e^{2\sqrt{\frac{N-1}{n(n+N-1)}}\phi}d\Omega_n^2\label{3.20a}
\ee
is an $(N+n+1)$--dimensional vacuum metric invariant under the group $SO(n+1)$.  Here $d\Omega_n^2$ is  the standard metric of the $n$-dimensional sphere. 

A particular solution of conditions (\ref{3.17a})--(\ref{3.24a}), for any $N>1$,  is given by
\be
\gamma_{ij}={\text{diag}}(+1,-1,-1,..)\ ,\ \ P_{ij}=P_{ji}=const\ ,\ \ P^i_{\ i}=0\ ,\ \lambda=0.
\label{3.25a}
\ee
For $N=2$ conditions (\ref{3.17a})--(\ref{3.24a}) can be solved in full generality. They lead either to (\ref{3.25a}) or to $c=0$ and to $\gamma e^{P\tau}$ equivalent  to the metric 
\be
(\gamma e^{P\tau})_{ij}dx^idx^j=\frac{dudv}{(1+\frac{\lambda}{4} uv)^2}+\tau h(u)du^2\ ,\label{3.35}
\ee
where $h$ is an arbitrary function of coordinate $u$. In the next section we find solutions of  equations (\ref{3.14}), (\ref{3.23}) and construct  corresponding vacuum metrics. In section 3 we discuss  properties of these metrics.

\section{Multi dimensional vacuum  metrics}
In \cite{1} we gave  examples of vacuum metrics derived by our method. 
Other solutions with  $n>1$ can be obtained by inspection of the Gross-Perry metrics \cite{GP}. Let  $\lambda=0$ and $s=s(r)$ be a function of  a new coordinate $r$.  
Then equations (\ref{3.14}), (\ref{3.23}), (\ref{3.21}) take the form
\be
\left(\frac{\beta \phi'}{\alpha}\right)'=\alpha\beta V_{,\phi}\label{3.26}
\ee
\be 
(1-\frac{1}{N})\frac{\beta'^2}{\alpha^2}-\frac{\beta^2\phi'^2}{\alpha^2}+2\beta^2V=2c
\ ,\label{3.27}
\ee
\be
\tau'=\frac{2\alpha}{\beta}\ ,\label{3.35b}
\ee
where the prime denotes the derivative with respect to $r$ and $\alpha=s'$. 
Metric (\ref{3.4}) is given by
\be
\tilde g=-\alpha^2dr^2+\beta^{2/N}(\gamma e^{P\tau})_{ij}dx^idx^j\ .\label{3.35a}
\ee

 If $N=n=2$ equations  (\ref{3.26}), (\ref{3.27}) are satisfied by functions $\alpha$, $\beta$, $\phi$ corresponding to the
Gross-Perry metric \cite{GP}. Changing parameters in these functions leads to the
following solutions for arbitrary dimensions $N>1$ and $n>1$
\be
\alpha=\alpha_0|r|^{-l-1}|r-r_0|^{l-p}|r+r_0|^{l+p}\label{3.28}
\ee
\be
\beta=\beta_0(r^2-r_0^2)\alpha\label{3.29}
\ee
\be
e^{\sqrt{\frac{n+N-1}{n\left(N-1\right)}}\phi}=(n-1)|r\alpha|\ .\label{3.30}
\ee
Here $l$ is a number defined by $n$ and $N$
\be
l=\frac{n+N-1}{(n-1)(N-1)}\label{3.31}
\ee
and $p$,   $\alpha_0$, $\beta_0$ and $r_0\neq 0$ are parameters related to the
constant $c$ via
\be
c=2\beta_0^2r_0^2\left[\frac{n}{n-1}-p^2\frac{(n-1)(N-1)^2}{N(n+N-1)}\right]\ .\label{3.32}
\ee
Integrating equation (\ref{3.35b}) yields 
\begin{equation}
\tau= \frac{1}{\beta_0 r_0}\ln{\left|\frac{r+r_0}{r-r_0}\right|}+\tau_0\ .\label{3.32a}
\end{equation}
Due to a
freedom of
transformations of $r$, $P$ and $\gamma$ we can assume 
\be
r_0>0\ ,\ \ |\alpha_0|=\frac{1}{n-1}\ ,\ \ \beta_0=1\ ,\ \ \tau_0=0\label{3.32b}
\ee
(note that a sign of $\alpha_0$ can be still 
adjusted to have  $\beta>0$ for $r\neq 0,\pm r_0$). Thus, $p$ and $r_0>0$ remain as free parameters.

Let $N=2$. In the case (\ref{3.25a})  and $c>0$ the matrix $P$ can be diagonalized by a 2-dimensional Lorentz transformation. Hence, one obtains
\be
\tilde g=-\alpha^2dr^2+\beta\left(e^{\pm\tau\sqrt{c}}dt^2-e^{\mp\tau\sqrt{c}}
 dy^2\right)\ ,\label{3.33}
 \ee
 where $t$ and $y$ denote coordinates $x^i$. Substituting
 (\ref{3.28})-(\ref{3.32b}) into (\ref{3.33}) and (\ref{3.20a})  yields 
the following (n+3)-dimensional vacuum metric
\be
g=\left|\frac{r-r_0}{r+r_0}\right
|^{p'-q}dt^2-\left|\frac{r-r_0}{r+r_0}\right|^{p'+q}dy^2-
\frac{|r+r_0|^{\frac{2p'+2}{n-1}}}{|r|^{\frac{2n}{n-1}}|r-r_0|^{\frac{2p'-2}{n-1}}}\left(\frac{dr^2}{(n-1)^2}+r^2d\Omega_n^2\right)\
.\label{3.41}
\ee
Parameters $p'$ and $q$ are related to $p$ and $c$ by
\be
p'=\frac{n-1}{n+1}p\ ,\ \ q=\pm\frac{\sqrt{|c|}}{r_0}\ .\label{3.41a}
\ee
 Because of
(\ref{3.32}) they are constrained by 
\be
(n+1)p'^2+(n-1)q^2=2n\ .\label{3.33a}
\ee
For n=2 solution (\ref{3.41}) is exactly the Gross-Perry metric \cite{GP} under the identification
\be
r_0=m,\quad
p'=\frac{1}{\alpha}(\beta+1),\quad
q=\frac{1}{\alpha}(\beta-1).
\ee
Here $m,\ \alpha$ and $\beta$ are parameters used by Gross and Perry,
constrained  by the condition $\alpha=\sqrt{\beta^2+\beta+1}$.

If $c<0$ the matrix $P$ can be put into the off diagonal form. Instead of (\ref{3.33}) one obtains 
\be
\tilde
g=-\alpha^2dr^2+\beta\left[\cos(\tau\sqrt{|c|})(dt^2-dy^2)\pm 2\sin{(\tau\sqrt{|c|})}
  dtdy\right]\ .\label{3.40a}
\ee
In this case the vacuum metric corresponding to (\ref{3.28})-(\ref{3.32b})  reads
\beq\nonumber
g&=&\left|\frac{r-r_0}{r+r_0}\right|^{p'}\left[\cos{\left(q\ln{\left|\frac{r+r_0}{r-r_0}\right|}\right)}(dt^2-dy^2)+
2\sin{\left(q\ln{\left|\frac{r+r_0}{r-r_0}\right|}\right)}
  dtdy\right]\\ &&-
\frac{|r+r_0|^{\frac{2p'+2}{n-1}}}{|r|^{\frac{2n}{n-1}}|r-r_0|^{\frac{2p'-2}{n-1}}}\left(\frac{dr^2}{(n-1)^2}+r^2d\Omega_n^2\right)\
.\label{3.44}
\eeq
Relation (\ref{3.41a}) is still valid, but now parameters $p'$, $q$ are
  constrained  by
\be
(n+1)p'^2-(n-1)q^2=2n\ .\label{3.44a}
\ee

If $N=2$ and 
\be 
p=\pm\frac{\sqrt{2n(n+1)}}{n-1}\label{3.45}
\ee
then it follows from (\ref{3.32}) that $c=0$ and one can merge solutions (\ref{3.28})-(\ref{3.30}) with 
metric (\ref{3.35}) for $\lambda=0$. In this way the following vacuum metric is obtained
\beq\label{3.47}
g&=&\left|\frac{r-r_0}{r+r_0}\right|^{\pm\sqrt{\frac{2n}{n+1}}}\left(dudv+\ln{\left|\frac{r+r_0}{r-r_0}\right|}
  h(u)du^2\right)\\\nonumber
&&-\frac{|r+r_0|^{\frac{2}{n-1}(\pm \sqrt{\frac{2n}{n+1}}+1)}}{|r|^{\frac{2n}{n-1}}|r-r_0|^{\frac{2}{n-1}(\pm \sqrt{\frac{2n}{n+1}}-1)}}\left(\frac{dr^2}{(n-1)^2}+r^2d\Omega_n^2\right)\ .
\eeq
In the case $n=2$, $h(u)=0$   metric (\ref{3.47}) with the lower sign coincides with the metric given by 
Millward \cite{Mil} under the identification
\be b=\frac{1}{\sqrt{3}}\ln{\left|\frac{r-r_0}{r+r_0}\right|}\ ,\ M=\frac{\sqrt{3}}{2}r_0\ .\ee

For $N>2$ one can easily construct vacuum solutions based on relations (\ref{3.20a}), (\ref{3.25a}),  (\ref{3.35a})-(\ref{3.32b}). They  generalize metrics (\ref{3.41}) and (\ref{3.44}). In this case a classification of symmetric tensors (here $P_{ij}$) in multidimensional Lorentzian manifolds \cite{CMPP} can be useful  in order to distinguish nonequivalent solutions. One can also construct metrics which generalize (\ref{3.47}) by taking  $\gamma e^{P\tau}$ corresponding to the metric
\be
dudv+\tau h(u)du^2+\sum_{a=1}^{N-2}{ e^{c_a\tau}dy^2_a}\ ,\label{3.48}
\ee
where constants $c_a$ are constrained by
\be
\sum_{a=1}^{N-2}c_a=0. \label{3.48a}
\ee
In this case we can use functions defined by (\ref{3.28})-(\ref{3.32b}) with constant $c$ given by
\be
c=\frac{1}{2}\sum_{a=1}^{N-2}c_a^2\ .\label{3.49}
\ee

\section{Discussion}
In adition to $SO(n+1)$ symmetries metrics (\ref{3.41}) and (\ref{3.44})
admit one timelike and one spacelike  Killing vector (note that interpretation
of $\partial_t$ and $\partial_y$ in case (\ref{3.44}) can change depending on
value of $r$). Metric (\ref{3.41}) is static and metric (\ref{3.44}) is stationary. In the limit $r\rightarrow \infty$ they behave like
\be\label{4.4}
dt^2-dy^2-r^{-2\frac{n-2}{n-1}}\left(\frac{dr^2}{(n-1)^2}+r^2d\Omega^2_n\right).
\ee
Under the change 
$
r'=r^{\frac{1}{n-1}}
$
metric (\ref{4.4}) takes the standard  form of the (n+3)-dimensional Minkowski metric. Thus, metrics (\ref{3.41}) and
(\ref{3.44}) are asymptotically flat on surfaces $y=const$. 
 
Metric (\ref{3.47}) has a null Killing vector field $\partial_v$ and it
belongs to generalized Kundt's class \cite{C}. If $r\rightarrow \infty$ it
tends asymptotically to the flat metric in the form
\be
dudv-r^{-2\frac{n-2}{n-1}}\left(\frac{dr^2}{(n-1)^2}+r^2d\Omega^2_n\right).
\ee

Generalizing results of \cite{CP} for the Gross-Perry metric to arbitrary
n, one can show that both metrics (\ref{3.41}) and (\ref{3.44}) are of algebraic
type $I$.
For $h\neq 0$ metric  (\ref{3.47}) is of algebraic type $II_i$ 
and for $h(u)=0$ it is of type D. Aligned null vector fields for metrics   (\ref{3.41}),
(\ref{3.44}) and (\ref{3.47}) are given in Appendix A.

All metrics \ref{3.41}),
(\ref{3.44}) and (\ref{3.47}) are singular at $r=\pm r_0$ and $r=0$. Near $r=0$ they behave as 
\be\label{3.38?}
dt^2-dy^2-r^{-\frac{2n}{n-1}}\left(\frac{dr^2}{(n-1)^2}+r^2d\Omega^2_n\right).
\ee
Substituting $r'=r^{-\frac{1}{n-1}}$ shows that (\ref{3.38?}) is the flat metric. Thus, $r=0$ is a coordinate singularity.
By calculating the Kretschmann invariant (see Appendix B) it can be shown, that singularity at $r=\pm r_0$ is essential for all values of parameters in the case of metrics (\ref{3.41}) and (\ref{3.47}). In the case of metric (\ref{3.44}) the singularity at $r=r_0$ and $r=-r_0$ is essential when, respectively, $p'<n$ or $p'>-n$. 
For $p'>n$ or $p'<-n$ the geodesic distance along $\partial_r$ tends to infinity when $r\rightarrow r_0$ or $r\rightarrow -r_0$, respectively. Thus, these regions represents an infinity different from that given by $r\rightarrow \infty$. For these values of parameters the Riemann tensor (in an orthonormal basis) tends to zero when $r\rightarrow r_0$ or $r\rightarrow -r_0$, respectively. However, the asymptotic metric is not  the (n+3)-dimensional Minkowski metric. Its coefficients in front of $dt$ and $dy$ tend to zero whereas the coefficient in front of $d\Omega_n^2$ tends to infinity like the geodesic distance to the power $2(p'-1)/(p'-n)$ or $2(p'+1)/(p'+n)$, respectively.

Since  metrics (\ref{3.41}),
(\ref{3.44}) are invariant under the nonnull field $\partial_u$ they can be interpreted in the context of the  Kaluza-Klein theory. Then metric (\ref{3.41}) is equivalent  to the scalar field given by $g_{yy}$ and the asymptotically flat (n+2)-dimensional  metric induced on the surface y=const. The case $n=2$ (the Gross-Perry metric) was studied in this framework by Ponce de Leon \cite{PL}. In order to interpret metric (\ref{3.44}) with $q\neq 0$  in this vein   one can write it in the form
\be
g=-\Phi(dy-A_0dt)^2+g_{n+2}\ .\label{4.10}
\ee
Here
\be
A_0=\tan{\left(q\ln{\left|\frac{r+r_0}{r-r_0}\right|}\right)}\label{4.11}
\ee
is the electromagnetic potential,
\be
\Phi=\left|\frac{r-r_0}{r+r_0}\right|^{p'}\cos{\left(q\ln{\left|\frac{r+r_0}{r-r_0}\right|}\right)}\label{4.12}
\ee
corresponds to a scalar field and
\be
g_{n+2}=\left|\frac{r-r_0}{r+r_0}\right|^{p'}\frac{dt^2}{\cos{\left(q\ln{\left|\frac{r+r_0}{r-r_0}\right|}\right)}}-
\frac{|r+r_0|^{\frac{2p'+2}{n-1}}}{|r|^{\frac{2n}{n-1}}|r-r_0|^{\frac{2p'-2}{n-1}}}\left(\frac{dr^2}{(n-1)^2}+r^2d\Omega_n^2\right)\label{4.13}
\ee
defines, modulo a power of $\Phi$,  a (n+2)-dimensional  metric. This metric is Lorentzian and asymptotically flat for large values of $r$ and becomes singular when $r$ diminishes to a value satisfying condition  $q\ln{\left|\frac{r+r_0}{r-r_0}\right|}=\pm \pi/2$.

\noindent 
{\bf Acknowledgements}. 
This work was partially supported by the Polish  Committee for Scientific Research (grant 1 PO3B 075 29). 

\newcounter{appendixc}\refstepcounter{appendixc}
\setcounter{equation}{0}
\renewcommand{\theequation}{\Alph{appendixc}.\arabic{equation}}
\appendix   
\section*{Appendix A}
The aligned null direction  is given by
\beq\nonumber
&\hat{l}&=\left(\left|\frac{r-r_0}{r+r_0}\right|^{p'-q}+\frac{1}{2}f^2\right)dt+\left(\left|\frac{r-r_0}{r+r_0}\right|^{p'}-\frac{1}{2}f^2\left|\frac{r-r_0}{r+r_0}\right|^{q}\right)dy+\\&&+f\frac{|r+r_0|^{\frac{p'+1}{n-1}}}{(n-1)|r|^{\frac{n}{n-1}}|r-r_0|^{\frac{p'-1}{n-1}}}dr
\eeq
for metric (\ref{3.41})
and by 
\beq\nonumber
&\hat{l}&=\left(\sin{\left(q\ln{|\frac{r+r_0}{r-r_0}|}\right)}+1\right)\left(1-\frac{1}{2}f^2\left|\frac{r-r_0}{r+r_0}\right|^{p'}\frac{\cos{\left(q\ln{|\frac{r+r_0}{r-r_0}|}\right)}}{\left(\sin{\left(q\ln{|\frac{r+r_0}{r-r_0}|}\right)}+1\right)^2}\right)dt+\\&&+\left(\frac{1}{2}f^2\left|\frac{r-r_0}{r+r_0}\right|^{p'}-\cos{\left(q\ln{|\frac{r+r_0}{r-r_0}|}\right)}\right)dy+f\frac{|r+r_0|^{\frac{p'+1}{n-1}}}{(n-1)|r|^{\frac{n}{n-1}}|r-r_0|^{\frac{p'-1}{n-1}}}dr
\eeq
for  metric (\ref{3.44}).
The function f is a solution of the polynomial equation 
\be
f^4-8f^2\left(\frac{p'}{q}+\frac{2q(n-1)rr_0}{n(r^2+r_0^2-2p'rr_0)}\right)A-16A^2=0,
\ee
where
\be
A=\left|\frac{r-r_0}{r+r_0}\right|^{p'-q}
\ee
in the case (\ref{3.41}) and
\be
A=\left|\frac{r-r_0}{r+r_0}\right|^{-p'}\left(\sin{\left(q\ln{|\frac{r+r_0}{r-r_0}|}\right)}+1\right)
\ee
in the case (\ref{3.44}).
For  metric (\ref{3.47}) the aligned null directions are defined by
\be
\hat{n}=\left|\frac{r+r_0}{r-r_0}\right|^{\pm\sqrt{\frac{2n}{n+1}}}du
\ee
and
\be
\hat{l}=\frac{1}{2}\left(\ln{\left|\frac{r-r_0}{r+r_0}\right|}+f^2\left|\frac{r+r_0}{r-r_0}\right|^{\pm\sqrt{\frac{2n}{n+1}}}\right)du+\frac{1}{2}dv+\frac{|r+r_0|^{\frac{1}{n-1}(\pm \sqrt{\frac{2n}{n+1}}+1)}}{(n-1)|r|^{\frac{n}{n-1}}|r-r_0|^{\frac{1}{n-1}(\pm \sqrt{\frac{2n}{n+1}}-1)}}dy.
\ee
In this case
\be
f^2=\left(\mp\sqrt{\frac{n+1}{2n}}+\frac{2(n^2-1)rr_0}{n(n+1)(r^2+r_0^2)\mp\sqrt{2n(n+1)}rr_0}\right)h(u)\left|\frac{r+r_0}{r-r_0}\right|^{\mp\sqrt{\frac{2n}{n+1}}}.
\ee
\appendix
\section*{Appendix B}
\setcounter{equation}{0}
\renewcommand{\theequation}{B.\arabic{equation}}
The Kretschmann invariant for metrics (\ref{3.41}) and (\ref{3.44}) has the following form
{\setlength\arraycolsep{2pt}
\beq\nonumber
R_{\mu\nu\delta\sigma}R^{\mu\nu\delta\sigma}&=&16n(n-1)r_0^2\left|r\right|^{\frac{2(n+1)}{n-1}}\left|r-r_0\right|^{-\frac{4(n-p')}{n-1}}\left|r+r_0\right|^{-\frac{4(n+p')}{n-1}}\left((n^2+n-2p'^2)r^4\right.+\\\nonumber&&\left.+2p'(n(n+1)(p'^2-3)+2(1+p'^2))r^3r_0\right.+\\\nonumber&&\left.-(4-3n-5n^2-2(-2+n(n+3))p'^2+(4+n(n+3))p'^4)r^2r_0^2\right.+\\&&\left.+2p'(n(n+1)(p'^2-3)+2(1+p'^2))rr_0^3+(n^2+n-2p'^2)r_0^4\right)\ .\label{A1}
\eeq}
In the case of metric (\ref{3.47}) the Kretschmann invariant is given by (\ref{A1}) with $p'=\pm\sqrt{\frac{2n}{n+1}}$.

\end{document}